\documentclass[12pt]{article}        

\usepackage{amsmath}
\usepackage{amssymb}
\usepackage{epsfig}

\usepackage{cite}

\usepackage{epic}
\usepackage{fancybox}

\newcommand{\umf}{\mathfrak{u}}

\begin{document}                     

\allowdisplaybreaks

\begin{titlepage}


\begin{flushright}
CERN-TH-98-235\\
UTHEP-98-0701
\end{flushright}

\vspace{1mm}
\begin{center}
{\Large\bf
  Global Positioning of Spin GPS Scheme \\
  for Half-Spin Massive Spinors
}
\end{center}
\vspace{3mm}

\begin{center}
{\bf S. Jadach$^{a,b}$},
{\bf B.F.L. Ward$^{a,c,d}$}
{\em and}
{\bf Z. W\c as$^{a,b}$} \\

\vspace{1mm}
{\em $^a$CERN, Theory Division, CH-1211 Geneva 23, Switzerland,}\\
{\em $^b$Institute of Nuclear Physics,
  ul. Kawiory 26a, Krak\'ow, Poland,}\\
{\em $^c$Department of Physics and Astronomy,\\
  The University of Tennessee, Knoxville, TN 37996-1200, USA,\\
  $^d$SLAC, Stanford University, Stanford, CA 94309, USA} 
\end{center}

\vspace{10mm}
\begin{abstract}
We present a simple and flexible method of keeping track of the complex 
phases and spin quantization axes for half-spin initial- and final-state
Weyl spinors in scattering amplitudes of Standard Model high energy physics 
processes. Both cases of massless and massive spinors are discussed.
The method is demonstrated and checked numerically for spin correlations in 
$\tau \bar\tau$ production and decay.
Its main application will be in the forthcoming work of combining effects due to 
multiple photon emission (exponentiation) and  spin, embodied in the 
Monte Carlo event generators for production and decay of unstable fermions
such as the $\tau$ lepton, t-quark and hypothetical new heavy particles.
\end{abstract}

\vspace{1mm}
\begin{flushleft}
{\bf CERN-TH/98-235\\
  UTHEP-98-0701\\
     July~1998}
\end{flushleft}

\footnoterule
\noindent
{\footnotesize
\begin{itemize}
\item[${\dagger}$]
Work supported in part by 
the US DoE contracts DE-FG05-91ER40627 and DE-AC03-76SF00515,
Polish Government grants 
KBN 2P03B08414, 
KBN 2P03B14715, 
Maria Sk\l{}odowska-Curie Joint Fund II PAA/DOE-97-316,
and Polish-French Collaboration within IN2P3.
\end{itemize}
}

\end{titlepage}

\section{\normalsize Introduction: What is the problem?}
Weyl spinors (Ws) are used in very powerful algorithms to calculate 
multiple-particle and in particular multiple-photon spin amplitudes 
\cite{Gastmans-book,CALKUL}.
In Ws calculations the  complete  control of the spin-dependent complex phase, 
i.e. of spin quantization frames, is usually neglected. This is not a problem,
when one is interested in cross sections for stable particles, but it becomes
a crucial limitation if one needs to combine them with the spin amplitudes 
for the decay of an almost stable (narrow resonance) fermion. 

The complete spin amplitude for production and decay is given by the formula:
\begin{equation}
  \label{combine-spin-amplitudes}
  {\cal M} 
  = \sum_{\lambda=\pm 1/2} {\cal M}^{\rm prod.}_\lambda
                           {\cal M}^{\rm decay}_\lambda,
\end{equation}
where $\lambda$ is the projection of the {\em intrinsic} spin on the $z$-axis in 
a certain fermion rest frame 
(see discussion in the appendix of ref.~\cite{Wick:1962}, and also ref.~\cite{Wigner:1939}).
Let us call this particular fermion rest frame the spin quantization reference frame, SQRF.
When, in the above formula,
both production and decay spin amplitudes 
are obtained from Feynman rules in {\em the same} calculation, 
with the same intermediate fermion states $u(p,\lambda)$, 
the situation is simple --
we even do not need to know very precisely the SQRFs.
The above is not true in the many important practical cases such as
the production of the $\tau$ lepton,
t-quark and hypothetical supersymmetric particles.
The reason is the following:
the fermion may have many decay modes, with complicated multiparticle final states,
and complicated half-phenomenological decay matrix elements, calculated
and implemented separately. For instance in the case of $\tau$ decay,
the TAUOLA Monte Carlo \cite{tauola2.4} represents by itself an independent
computational project. In fact every one out of  more than twenty  decay 
modes of the $\tau$ represents an independent project both from the computational 
and from the theoretical perspective.

For these practical reasons,
combining production and decay at the {\em density matrix }
level, even if it seems at first more complicated,  is worth an effort, because 
formula ~(\ref{combine-spin-amplitudes}) would lead to independent sets of spin 
amplitudes for distinct combinations of decay modes 
of unstable final-state fermions.
The density matrix formalism allows for clean modularity in the calculations and in 
the architecture of the corresponding Monte Carlo event generators.

The price one has to pay (for spin effects  not to be destroyed) is that the 
complete combined density matrix 
for all final-state unstable particles must be provided by the production program
with the careful definition of the corresponding SQRFs. That requires, 
in particular, full control of the relative phases of all production spin 
amplitudes.

In our paper we shall rely on massless and massive Weyl spinors
defined in two papers of Kleiss and Stirling (KS) 
\cite{KleissStriling:1985,KleissStriling:1986}. 
An analogous set of massless spinors is also defined in Ref.~\cite{Beijing}.
By supplementing these techniques with rules for 
complete  control of spin quantization axes (complex phases) 
we turn them into a powerful tool for calculations of unstable half-spin fermions, 
useful for a Monte Carlo simulation of production and decay. 
The aim of the present work is to adapt the spin amplitude techniques to the
framework established in the works 
\cite{jadach_was:1984,KORALB:1985,tauola2.4}, see also \cite{Tsai:1971,Kuhn:1984}.
We shall use it in the forthcoming work on exclusive coherent exponentiation
\cite{CEEX}.

Let us start with the fundamental question of defining 
properly the spin quantization axes (and complex phases) for half-spin Weyl spinors.
We  present details of the lowest-order $\tau$-pair production process
\begin{equation}
  \label{tau-process}
  e^-(p_1,\lambda_1) +e^+(p_2,\lambda_2) \to \tau^-(q_1,\mu_1) +\tau^+(q_2,\mu_2)
\end{equation}
with spin amplitudes
${\cal M}^{\rm prod}_{\lambda_1 \lambda_2 \mu_1 \mu_2}$
where $\mu_1,\mu_2=\pm 1$ denote twice the spin $z$-projection
of the corresponding $\tau$ lepton in its rest frame,
and $\lambda_i$ are analogous spin indices of $e^\pm$.
To be clear, we will not necessarily always restrict ourselves to
$\lambda_i$ and $\mu_i$ being the conventional Jacob-Wick helicities \cite{JacobWick:1959}.

The full differential cross section for the combined $\tau$-pair production and decay
reads as follows
\begin{equation}
  d\sigma = d\sigma^{\rm prod}(q_1,q_2)\;
            {d\Gamma_1^{\rm decay} (\tau^- \to X_1) \over \Gamma_1^{\rm decay}} \; 
            {d\Gamma_2^{\rm decay} (\tau^+ \to X_2) \over \Gamma_2^{\rm decay}} \; 
            \left( \sum_{a,b=0}^3 
              \varepsilon_1^a \varepsilon_2^b R_{abcd}\; h_1^c h_2^d \right)
\end{equation}
where $d\sigma^{\rm prod}$ is the unpolarized differential cross section for 
the production
process only, and $d\Gamma_i^{\rm decay}$ are unpolarized differential partial
widths for decays of the $\tau^\pm$.
The {\em polarimeter} vectors $h_i^a$ are related to the $\tau$ decay matrix element
(see eq.~(2.18) of ref.~\cite{jadach:1985} for its precise definition)%
\footnote{ The vectors $h_i^a$ depend, in general, on momenta of all $\tau$ decay products.},
and $\varepsilon_i$ are conventional spin polarization vectors of $e^\pm$
parametrizing their spin density matrices 
$\rho_{\lambda, \bar \lambda} = (1+\vec{\sigma}\cdot\vec{\varepsilon})_{\lambda, \bar \lambda}/2$.

All spin effects are embodied in the last term, which includes 
the full spin density
matrix related to the production spin amplitudes as follows:
\begin{equation}
  \label{R-matrix}
  \begin{split}
    &R_{abcd} = N^{-1} \sum_{\lambda_i,\mu_i,\bar{\lambda}_i,\bar{\mu}_i }
             {\cal M}_{\lambda_1 \lambda_2 \mu_1 \mu_2}
             ({\cal M}_{\bar{\lambda}_1 \bar{\lambda}_2 \bar{\mu}_1 \bar{\mu}_2})^\star
             \sigma^a_{\lambda_1 \bar{\lambda}_1}
             \sigma^b_{\lambda_2 \bar{\lambda}_2}
             \sigma^c_{\bar{\mu}_1 \mu_1 }
             \sigma^d_{\bar{\mu}_2 \mu_2 },\\
    &N= \sum_{\lambda_i,\mu_i} |{\cal M}_{\lambda_1 \lambda_2 \mu_1 \mu_2}|^2,\quad
     a,b,c,d=0,1,2,3,
  \end{split}
\end{equation}
where $\sigma^k$ for $k=1,2,3$ are Pauli matrices and
$\sigma^0_{\lambda,\mu} = \delta_{\lambda,\mu}$ is just the unit matrix.

The translation  in  eq.~(\ref{R-matrix}) 
from spin amplitudes to $R_{abcd}$, and the analogous
translation from decay spin amplitudes to the vectors $h_c$ and $h_d$,
occurs in the SQRF of $\tau^-$ for indices  $c$, $\mu_1$, $\bar \mu_1$
and in the SQRF of $\tau^+$ for indices $d$, $\mu_2$, $\bar \mu_2$.
Let us stress again that this translation occurs not in some arbitrary 
$\tau^\pm$ rest frame but in {\em precisely the same} SQRF
of $\tau^\pm$, where $\mu_i$ was originally defined as a 
spin projection on the $z$-axis!
Not only the $z$-axis has to be known, but also the $x$- and 
$y$-axes\footnote{The change of SQRF
  by a rotation around the spin quantization $z$-axis by an angle 
  $\phi$ introduces the factor $\exp((i\phi/2)\mu_1)$ (rotation in spinor
  representation)
  in the spin amplitudes and consequently an 
  $R_3(\phi)$ rotation in normal space (i.e. adjoint to spinor 
  representation) acting on 
  the index $c$ in $R_{abcd}$. N.B. a simplest example of Wick rotation
  is actually $R_3(\phi)$.
  }.
Being in the SQRF makes it also legitimate to use Pauli matrices for the 
translation
of spin indices into vector indices in eq.~(\ref{R-matrix})
\footnote{ Pauli matrices play here the role of Clebsch-Gordan coefficients 
  to combine
  two half-spin representations of the rotation group  
  (two indices in the spin density matrix) 
  into spin 0 and 1 angular momentum objects,
  see eq.~(A.10) in ref.~\protect\cite{jadach:1985}.
  }.

In the practical calculations that implement the above methodology, 
see KORALB \cite{KORALB:1985} and KORALZ \cite{KORALZ_4.0},
a special {\em subprogram} {\tt TraLor} is available, which implements 
the Lorentz transformation from the SQRF\footnote{The {\tt TraLor}
{\em subprogram} actually provides explicit definition of the SQRF frame.} 
of each $\tau^\pm$ down to 
the laboratory system. 
It is used to transform the momenta of all $\tau^\pm$ decay products into the 
laboratory frame.
This arrangement fully guarantees that all spin effects, 
including spin correlations, are properly reproduced in the Monte Carlo event generation.

Having convinced the reader that the knowledge and the proper use of 
the SQRF for each
fermion is very important, let us look into how it is defined.
The obvious choice is to employ the standard definition of spin states by
Jacob and Wick (JW) \cite{JacobWick:1959}, the so-called helicity states.
In ref. \cite{jadach_was:1984} spinors of
outgoing $\tau^\pm$ (and also of $e^\pm$ beams) were defined 
using a variant of the JW prescription%
\footnote{The JW \cite{JacobWick:1959} recipe was modified 
  in ref. \protect\cite{jadach_was:1984}, and 
  in fact this choice is closer to that in ref.~\protect\cite{Wick:1962}.
  For final fermions the  $y$-axis was placed in the reaction plane, while
  in the JW prescription it is related by direct rotation 
  (around the axis perpendicular
  to the reaction plane) to the laboratory $y$-axis.}.
In the JW method, states with definite spin projection on the $z$-axis (helicity states)
are defined by construction in the SQRF; the full Lorentz
transformation down to the laboratory frame is used as an inherent 
construction element in the definition of the spin state.
In KORALB \cite{jadach_was:1984,KORALB:1985} 
the JW spin quantization principle was also successfully used
to define spin states of $e^\pm$ and $\tau^\pm$
in the presence of one photon.

On the other hand, looking at the calculations 
of refs.~\cite{jadach_was:1984,KORALB:1985} 
it is quite clear that the JW method is well suited
to lowest order $e^-e^+\to \tau^-\tau^+$,
but that it is rather complicated to deal with already in the case $e^-e^+\to \tau^-\tau^+\gamma$
since the orientation of SQRF systems 
depends on the photon direction even in the soft-photon limit.
One definitely needs a better method to go beyond the single-photon case.
The natural candidates are methods based on the Weyl spinors,
 which have provend to be very 
successful for the evaluation
of multiple photon emission spin amplitudes for massless and massive spinors.
Generally  Weyl spinors (Ws) have their own internal definition of the SQRF,
substantially different from that in the Jacob-Wick method.
For example in the KS technique~\cite{KleissStriling:1985,KleissStriling:1986}
the spin quantization axis in the rest frame
of the massive fermion is related to some auxiliary light-like
vector $k_0$ involved in the definition of the massive spinor.
The typical definition of the spinor in the KS method involves a single 
``brutal'' transformation  from the reference fermion state  
into the state of definite mass, spin and momentum rather 
than the explicit  ``soft'' rotation/boost method of the JW method.
The problem is that in the transformation method of the Ws/KS techniques 
the phase,
and therefore the position of the $x$- and $y$-axes, is lost or undefined.
This is not a great problem for most of the practical applications of the Ws/KS techniques,
such as the calculation of multiple photon emission spin amplitudes,
which are usually done with the ``modulo phase factor'' approach anyway.
This feature of Ws/KS methods inhibits, however, 
the use of this method for unstable fermions.
The aim of this paper is to
show how to restore the knowledge of the complex phase and the SQRF for
the KS massive spinors of refs.~\cite{KleissStriling:1985,KleissStriling:1986}.

In the following we shall repeat many definitions of
refs.~\cite{KleissStriling:1985,KleissStriling:1986}
in detail. This is made necessary by the fact the central question is the control
of the spinor phases, and even knowledge that this phase is zero 
(in certain cases) 
forms an important new result!
Our final answer for the problem of control of Weyl spinor phases and
all three axes of the SQRF is very simple -- the reader should
not be misled by its apparent simplicity, 
it is an important and new supplement of the existing Ws/KS methods.

\section{\normalsize Basics of the Weyl technique}
For Dirac spinors, wherever an explicit representation is needed, we
employ the Weyl representation of gamma matrices%
\begin{equation}
  \gamma^{0} =
  \begin{bmatrix} 
    0 & I \\ 
    I & 0 
  \end{bmatrix},\quad
  \gamma^{k} =
  \begin{bmatrix} 
        0      & -\sigma_k \\ 
     \sigma_k  &    0
  \end{bmatrix},\quad
  \gamma^{5} =
  \begin{bmatrix} 
    I & 0 \\ 
    0 & -I 
  \end{bmatrix}.\quad
\end{equation}
In the Weyl representation, spinors transform under rotations around
the $k$-th axis
and boosts along the $j$-th axis as follows \cite{ItzyksonZuber}:
\begin{equation}
  \label{spinor-transformations}
S(R_{k}(\phi)) = 
   \exp
   \left( 
     -{i\over 2}\; \phi
     \begin{bmatrix} 
       \sigma_{k}  &    0 \\ 
       0         &    \sigma_{k}
     \end{bmatrix}
   \right),\qquad
S(B_j(\chi)) = 
   \exp
   \left( 
      {1\over 2} \; \chi
     \begin{bmatrix} 
       \sigma_j  &    0 \\ 
       0         &   -\sigma_j
     \end{bmatrix}
   \right).\qquad
\end{equation}
We define eigenstates with definite {\em massless} four-momentum $p^2=0$
and chirality $\lambda=\pm 1$
\begin{equation}
  \label{define_eigenstates}
  \not\! p u_\lambda(p) = 0,\quad
  \omega_\lambda u_\lambda(p) = u_\lambda(p),\quad
  \omega_\lambda \equiv {1\over 2} (1+\lambda\gamma_5),
\end{equation}
with the normalization condition 
$u_\lambda(p) \bar{u}_\lambda(p) = \omega_\lambda \not\!p.$
The four basic {\em massless} spinors are defined in a certain primary
reference frame (PRF), as follows:
\begin{equation}
\label{basic-spinors}
\umf_{+}(\zeta_{\uparrow}) =
  \begin{bmatrix} 
    \sqrt{2} \\ 0 \\ 0 \\ 0
  \end{bmatrix},\quad
\umf_{+}(\zeta_{\downarrow}) =
  \begin{bmatrix} 
    0 \\ \sqrt{2} \\ 0 \\ 0
  \end{bmatrix},\quad
\umf_{-}(\zeta_{\uparrow}) =
  \begin{bmatrix} 
    0 \\ 0 \\ 0 \\-\sqrt{2}
  \end{bmatrix},\quad
\umf_{-}(\zeta_{\downarrow}) =
  \begin{bmatrix} 
    0 \\ 0 \\ \sqrt{2} \\ 0
  \end{bmatrix},
\end{equation}
and they correspond to particles flying along the $z$-axis:
      $\zeta_{\uparrow}   = (1,0,0,1)$ 
and   $\zeta_{\downarrow} = (1,0,0,-1)$.
It is important to remember that
the relative phases in the above set of basic vectors are constrained
uniquely by the following relations
\begin{equation}
  \label{phase-relations}
  \umf_\pm(\zeta_{\downarrow}) = S(R_2(\pi)) \umf_\pm(\zeta_{\uparrow}),\quad
  \umf_+(\zeta_{\uparrow})   =  -{\not\!\eta} \umf_-(\zeta_{\uparrow}),\quad
  \umf_+(\zeta_{\downarrow}) =   {\not\!\eta} \umf_-(\zeta_{\downarrow}),\quad
\end{equation}
where%
\footnote{Our $\zeta_{\downarrow}$ corresponds to the vector called $k_0$
  in ref.~\cite{KleissStriling:1985} and $\eta$ is denoted there by $k_1$.
  }
$\eta=\hat{e}_x = (0,1,0,0)$, $\eta^2=-1$, 
$\eta\cdot\zeta_{\uparrow}=0$ and $\eta\cdot\zeta_{\downarrow}=0$.

\section{\normalsize Massless spinors}
The arbitrary massless spinor of momentum $p$ and chirality $\lambda$
is generated according to the KS method \cite{KleissStriling:1985},
out of the two constant basic 
spinors of opposite chirality
(two out of four in eq. (\ref{basic-spinors}))
\begin{equation}
\label{def-massless}
  u_\lambda(p) 
     = {1\over \sqrt{ 2p\cdot \zeta}} \not\!p \umf_{-\lambda}(\zeta),\qquad
     \zeta \equiv \zeta_{\downarrow}.
\end{equation}
Written  explicitly, see also ref.~\cite{Beijing}, it looks as follows:
\begin{equation}
u_{+}(p) =
  \begin{bmatrix} 
    \sqrt{p^+} \\ \sqrt{p^-}e^{i\phi} \\ 0 \\ 0
  \end{bmatrix},\quad
u_{-}(p) =
  \begin{bmatrix} 
     0 \\ 0 \\ -\sqrt{p^-}e^{-i\phi} \\ \sqrt{p^+}
  \end{bmatrix},\quad 
p^\pm \equiv p^0 \pm p^3,\quad 
p^1 + i p^2 \equiv p_T e^{i\phi}.
\end{equation}
The above expressions lead directly to an explicit expression for the 
{\em inner product} of the two massless spinors
  \begin{equation}
    \begin{split}
      & s_{+}(p_1,p_2) \equiv \bar{u}_{+}(p_1) u_{-}(p_2)
      = \sqrt{p_1^- p_2^+} e^{-i\phi_1} 
      -\sqrt{p_1^+ p_2^-} e^{-i\phi_2},\\
      & s_{-}(p_1,p_2) \equiv \bar{u}_{-}(p_1) u_{+}(p_2) 
      = -( s_{+}(p_1,p_2))^*.\\
    \end{split}
  \end{equation}
It is understood that the above formula is evaluated using four-momenta in 
the PRF, see eq. (\ref{basic-spinors}).
The same formula can be obtained using the original KS expression%
\footnote{ 
  The analogous expression in eq.~(3.9) of ref.\cite{KleissStriling:1985}
  misses factor 2.
  For the Levi-Civita tensor we take  $\epsilon_{0123}=1$.}:
\begin{equation}
  s_+(p,q) = 2\; {1\over \sqrt{2p\zeta}}\; {1\over \sqrt{2q\zeta}}\; 
  \left[ (p\zeta)(q\eta) - (p\eta)(q\zeta) 
    -i\epsilon_{\mu\nu\rho\sigma} \zeta^\mu \eta^\nu p^\rho q^\sigma 
  \right],
\end{equation}
which is applicable  in any  reference frame, not only in the PRF.

\section{\normalsize Massive spinors}
Spinors for the massive particle with four momentum $p$ (with $p^2=m^2$)
can also be defined using a construction similar to that in the massless case:
\begin{equation}
\label{def-massive}
  u(p,\lambda) 
    = {1\over \sqrt{2p\cdot \zeta}}\; (\not\!p +m)\; \umf_{-\lambda}(\zeta),\qquad
  v(p,\lambda) 
    = {1\over \sqrt{2p\cdot \zeta}}\; (\not\!p -m)\; \umf_{ \lambda}(\zeta).
\end{equation}
As we see, in order to obtain the $v$-spinor for the antiparticle  we have to 
substitute
$m\to -m$ and $\lambda \to -\lambda$, and this is a general rule
in the following.
Apart from the Dirac equation 
$(\not\!p -m) u(p,\lambda) =0$ and 
$(\not\!p +m) v(p,\lambda) =0$
the above spinors obey the following normalization and completeness relations:
\begin{equation}
      u(p,\lambda)\; \bar{u}(p,\lambda) 
           = {1\over 2} (1+ \lambda \gamma_5 {\not\!s} )\; (\not\!p +m),\quad
      v(p,\lambda)\; \bar{v}(p,\lambda) 
           = {1\over 2} (1+ \lambda \gamma_5 {\not\!s} )\; (\not\!p -m),
\end{equation}
where
\begin{equation}
  \label{s-vector}
  s = {p\over m} - \zeta\; {m\over p\cdot \zeta},
  \quad s^2\equiv -1,
\end{equation}
is the {\em spin quantization vector}; 
in the fermion rest frame, it points opposite to $\zeta$,
i.e. $\vec{s} \sim -\vec{\zeta}$, or in other words the spin quantization axis
for every fermion is {\em guided} by the single and common massless vector $\zeta$.

The definition of eq.~(\ref{def-massive}) can be rewritten explicitly
in terms of massless spinors as follows
\begin{equation}
\label{def-massive2}
     u(p,\lambda)
        = u_\lambda(p_\zeta)    +{m\over \sqrt{2p\zeta}} \umf_{-\lambda}(\zeta),\quad
     v(p,\lambda)
        = u_{-\lambda}(p_\zeta) -{m\over \sqrt{2p\zeta}} \umf_{\lambda}(\zeta),
\end{equation}
where 
\begin{equation}
  p_\zeta =
  p - \zeta\; {m^2\over 2p\cdot \zeta}, \quad p_\zeta^2 =0
\end{equation}
is the light-cone projection of $p$ obtained with the help of the 
auxiliary vector $\zeta$.
Equation~(\ref{def-massive2}) immediately
provides us also with the explicit {\em inner product} for the massive spinors:
\begin{equation}
  \begin{split}
    & \bar{u}(p_1,\lambda_1)  u(p_2,\lambda_2)
                =S(p_1,m_1,\lambda_1,  p_2,m_2,\lambda_2) \\
    & \bar{u}(p_1,\lambda_1)  v(p_2,\lambda_2)
                =S(p_1,m_1,\lambda_1,  p_2,-m_2,-\lambda_2) \\
    & \bar{v}(p_1,\lambda_1)  u(p_2,\lambda_2)
                =S(p_1,-m_1,-\lambda_1,  p_2,m_2,\lambda_2) \\
    & \bar{v}(p_1,\lambda_1)  v(p_2,\lambda_2)
                =S(p_1,-m_1,-\lambda_1,  p_2,-m_2,-\lambda_2) \\
  \end{split}
\end{equation}
where 
\begin{equation}
\label{inner-massive}
  S(p_1,m_1,\lambda_1,  p_2,m_2,\lambda_2)
  = \delta_{\lambda_1,-\lambda_2} s_{\lambda_1}({p_1}_\zeta, {p_2}_\zeta)
   +\delta_{\lambda_1, \lambda_2} 
  \left(
     m_1 \sqrt{ {2\zeta p_2 \over 2\zeta p_1}  }
    +m_2 \sqrt{ {2\zeta p_1 \over 2\zeta p_2}  }\;
  \right).
\end{equation}

For the sake of the subsequent discussion,
let us also write the explicit basic massive spinors in the particular case when
the fermion rest frame coincides with the PRF 
(in which $\zeta=(1,0,0,-1)$):
\begin{equation}
\label{basic-spinors-massive}
u(p,+) =
  \begin{bmatrix} 
    1 \\ 0 \\ 1 \\ 0
  \end{bmatrix},\quad
u(p,-) =
  \begin{bmatrix} 
    0 \\ 1 \\ 0 \\ 1
  \end{bmatrix},\quad
v(p,+) =
  \begin{bmatrix} 
    0 \\ -1 \\ 0 \\ 1
  \end{bmatrix},\quad
v(p,-) =
  \begin{bmatrix} 
    1 \\ 0 \\ -1 \\ 0
  \end{bmatrix}.
\end{equation}
We have omitted the trivial normalization factor $\sqrt{m}$.
The convention for relative phases is inherited from the 
underlying massless spinors (\ref{phase-relations}).
In particular the important phase relations
\begin{equation}
  \label{phase-relation2}
  S(R_2(\pi))u(p,+) = u(p,-),\quad
  S(R_2(\pi))v(p,+) = v(p,-).
\end{equation}

For ultrarelativistic $p^0\to\infty$, or for an almost massless 
fermion $m\to 0$,
the difference between $p$ and $p_\zeta$ is negligible
and the massive spinor becomes identical with the corresponding massless spinor:
\begin{equation}
  \begin{split}
    &\lim_{m\to 0}\; u(p,\lambda) = \lim_{m\to 0}\; v(p,-\lambda) = u_\lambda(p_\zeta),\\
    &\lim_{m_i\to 0}\; S(p_1,m_1,\lambda_1,  p_2,m_2,\lambda_2) 
    = \delta_{\lambda_1,-\lambda_2} s_{\lambda_1}({p_1}_\zeta, {p_2}_\zeta)
  \end{split}
\end{equation}
The validity of the above limits is restricted by the condition $p\cdot\zeta \gg m$,
i.e. they fail in the special case of $p$ parallel to $\zeta$ and
such cases have to be treated one by one, with special care.
In such a case, in eq.~(\ref{def-massive2}),
the second term, proportional to mass, survives and dominates.
The same phenomenon occurs in the inner product of eq.~(\ref{inner-massive}).
Thus, we have to remember that the naive procedure of
taking the massless fermion limit by means of straightforward omission of
all terms proportional to mass can be dangerous and the limiting procedure has to be
done very carefully%
\footnote{On the other hand, 
  keeping mass terms does not necessarily mean correct results
  in numerical calculations, because of machine rounding errors.}.

\section{\normalsize Again the central question and the GPS answer}
We now come to the central question of interpreting the massive fermion spinors
of eq.~(\ref{def-massive}), the spin vector of eq.~(\ref{s-vector}),
and finding out what the SQRF for these spinors is.
The states $u(p,\pm)$ and $v(p,\pm)$ represent {\em pure quantum mechanical} states, which
we intend to use in the Feynman rules to calculate scattering matrix elements.
As we see for instance from eq.~(\ref{s-vector}), $u(p,+)$ and $u(p,-)$ have
spin projections $+1/2$ and $-1/2$ on the vector $\vec{s}$ in the fermion rest frame.
Thus the vector $\vec{s}$, which is {\em guided} by $\zeta$, 
determines the $z$-axis of the SQRF.
This is the easy part of determining the SQRF.
The more difficult question is: 
Where are the $x$- and $y$-axes of the SQRF
corresponding to $u(p,+)$ and $u(p,-)$ defined in eq.~(\ref{def-massive})?
Is this question meaningful at all? Yes, it is meaningful because a
rotation $R_3(\phi)$ around the $z$-axis introduces a
$\lambda$-dependent phase in $u'(p,\lambda)$.
In fact one possible method of finding the $x$- and $y$-axes of the SQRF is
to inspect $u(p,\pm)$ in the rest frame of the fermion and to
adjust $\phi$ such that the relative phase of $u'(p,+)$ and $u'(p,-)$ is zero,
i.e. such that the relation  
$S(R_2(\pi))u'(p,+) = u'(p,-)$ holds.
This rule is, however, not very practical --
we would like to have at our disposal more practical, easy to apply, rules,
which determine the position of the same $x$- and $y$-axes
of the SQRF for both ($\lambda=\pm$) spinors defined
in eq.~(\ref{def-massive}).

The rules for determining all three spin quantization axes 
for $u(p,\pm)$ and $v(p,\pm)$ of eq.~(\ref{def-massive})
are the following:
\begin{itemize}
\item 
In the rest frame of the fermion, take the $z$-axis along $-\vec{\zeta}$.
\item
Place the $x$-axis in the plane defined by the $z$-axis from the previous point
and the vector $\vec{\eta}$, in the same half-plane as $\vec{\eta}$.
\item
With the $y$-axis, complete the right-handed system of coordinates. 
\end{itemize}
We call the above rules the Global Positioning of Spin scheme or in short the GPS rules,
and we shall call the GPS frame the SQRF determined by the GPS rules.
We give here the formal proofs of the above rules.

{\em Proof:} 
Since all massive spinors are defined in terms of two basic massless spinors 
$\umf_{+}(\zeta_{\downarrow})$ and $\umf_{-}(\zeta_{\downarrow})$,
it is sufficient to consider underlying massless spinors 
and to prove that after transforming them
from the PRF to the GPS frame they look the same
as in eq.~(\ref{basic-spinors}), up to a common phase factor.
In other words the two basic massless spinors 
$\umf_{+}(\zeta_{\downarrow})$ and $\umf_{-}(\zeta_{\downarrow})$
get {\em regenerated} in the GPS frame.
Without actually doing explicitly the transformation from the PRF to 
the GPS frame
we know that because of the Dirac equation, 
chirality deefinition (see eq. (\ref{define_eigenstates})\;),
and because $\zeta \equiv \zeta_{\downarrow} ={\rm const}\; (1,0,0,-1)$
the resulting spinors have to look as follows
\begin{equation}
\umf_{+}(\zeta_{\downarrow}) =
  \begin{bmatrix} 
    0 \\ c_2 \\ 0 \\ 0
  \end{bmatrix},\quad
\umf_{-}(\zeta_{\downarrow}) =
  \begin{bmatrix} 
    0 \\ 0 \\ c_3 \\ 0
  \end{bmatrix},
\end{equation}
where $c_2$ and $c_3$ are unknown complex constants.
We shall now exploit two facts: 
(a) in the local GPS frame, $\vec{\eta}$ lies
in $z$-$x$ plane (with positive $x$-component) 
and (b) the four-vector $\eta$ obeys
$\eta\cdot\zeta_{\downarrow}=0$ and $\eta^2=-1$.
The above implies that
$\eta = (-a,1,0,a),$ where $a$ is  a real constant, and therefore
\begin{equation}
  \not\!\eta =
  \begin{bmatrix} 
     0   &  0  &  0  &  1  \\ 
     0   &  0  &  1  & -2a \\ 
    -2a  & -1  &  0  &  0  \\ 
    -1   &  0  &  0  &  0
  \end{bmatrix}.
\end{equation}
The above completes the proof because we may exploit the relation
\begin{equation}
  \umf_{+}(\zeta_{\downarrow}) 
  = \not\!\eta\; \umf_{-}(\zeta_{\downarrow})
  = \begin{bmatrix} 
       0 \\ c_3 \\ 0 \\ 0
    \end{bmatrix},\quad
\end{equation}
see eq.~(\ref{phase-relations}), finding out that $c_3=c_2$, and therefore
we do really get back the same looking basic spinors in the GPS frame
as in the original PRF,
up to a spin-independent phase factor $c=c_3=c_2=r\exp(i\phi)$.
In fact one may advance even one step further and prove that the actual 
overall phase $\phi$ is exactly zero. 
We sketch a formal proof of this conjecture in the following.

{\em Proof:} 
Let us observe that the spinor transformations 
$S(B_1)$, $S(B_3)$ and $S(R_2)$,
i.e. transformations induced by boosts in the $x$-$z$ plane, 
and the rotation around the $y$-axis are purely real,
while the $S(R_3)$ induced by the rotation around the $z$-axis is 
complex and diagonal,
see eqs.~(\ref{spinor-transformations}).
We shall now analyse the complete string of the Lorentz transformations connecting
the PRF and the GPS frames of a given fermion with momentum $p$.
The first rotation $R_3(\phi_p)$ puts $\vec{p}$ in the $x$-$z$ plane
and introduces the following phases:
$\umf_{\lambda}(\zeta_{\downarrow}) \to \exp(i\phi_p/2\lambda)\umf_{\lambda}(\zeta_{\downarrow})$.
Next, the rotation $R_2$ and boost $B_3$ and again the rotation $R_2$
transform us to the frame where $p=(m,0,0,0)$ and $\zeta_{\downarrow}=(1,0,0,-1)$.
Since the corresponding transformations of spinors are real, 
we know that the resulting final $\umf_{\lambda}(\zeta_{\downarrow})$ 
preserves its complex phase!
We are not yet in the GPS frame --
the last transformation needed is the rotation $R_3(\phi')$ 
which brings $\vec{\eta}$ into the $x$-$z$ plane. 
The resulting net phase is $\exp((i/2)(\phi_p+\phi')\lambda)$.
Since we already know from the former proof of the GPS rules
that there is no $\lambda$ dependence in the total phase, 
the only possible solution is that the total change 
of the spinor phases is zero, $\phi_p+\phi'=0$, modulo $4n\pi$ for some
integer $n$, so that $\exp((i/2)(\phi_p+\phi')\lambda)=1$ as desired.
(It is also  possible to give the argument, 
based on geometry, that $\phi_p=-\phi'$.)
We have also checked numerically, using spinor transformations of 
eq.~(\ref{spinor-transformations}),
that the series of the above Lorentz transformation really regenerates 
the {\em basic}
spinors $\umf_\pm(\zeta)$ in the GPS frame.
For massive spinors, we have checked numerically that the massive spinors defined
in eq.~(\ref{def-massive2}) in the laboratory frame, 
when transformed properly to the GPS frame,
become identical to those of eq.~(\ref{basic-spinors-massive})!

Summarizing: the important property of the GPS rules is that the basic spinors
defined in eq.~(\ref{def-massive}), when transformed to the GPS rest frame
of the fermion, obtain  the same form (are regenerated), 
up to a real spin-independent, normalization constant, as the original ones,
defined in eq.~(\ref{basic-spinors-massive}).
The Lorentz transformation from any Lorentz frame, 
for instance from the laboratory reference system,
to the SQRF/GPS frame of every massive fermion is therefore uniquely defined.
Thanks to the GPS rules, we are finally in a situation 
as comfortable as in the Jacob-Wick method.
The  difference is however clear: 
in the JW method, the transformation from the laboratory frame to the SQRF
is the essence and the starting point of the method --
it is tied up to the fermion four-momentum in the laboratory system, 
and the axes of that frame.
In the GPS method the SQRF/GPS frame and
the Lorentz transformation from the laboratory to the SQRF/GPS frame 
are deduced with the help
of the ``backward engineering'' procedure guided by
the auxiliary vectors $\zeta$ and $\eta$ (common to all fermions),
without any direct reference to the laboratory system.

\begin{figure}[!th]
\centering
\setlength{\unitlength}{0.1mm}
\begin{picture}(1600,1200)
\put( -20, 0){\makebox(0,0)[lb]{
\epsfig{file=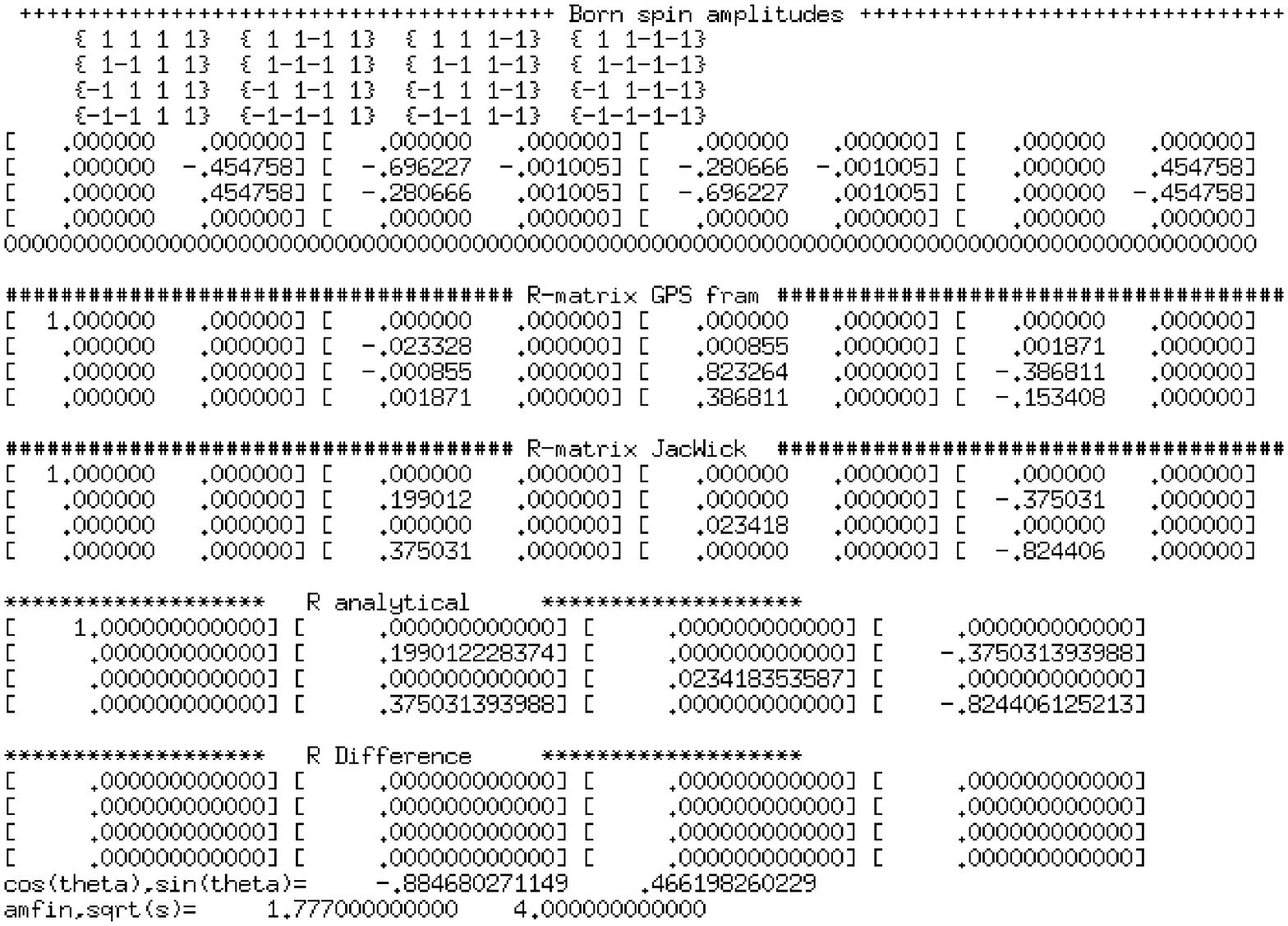,width=160mm,height=120mm}
}}
\end{picture}
\caption{\small\sf
  Numerical illustration of Wick rotation for one $\tau$-pair production event at 
  $\protect\sqrt{s}=$4 GeV, $m_\tau=1.777$ GeV, $\cos\theta=-0.884680271149$.
  In the upper part we print the complex spin amplitudes
  ${\cal M}^{\rm prod}_{\lambda_1 \lambda_2 \mu_1 \mu_2}$, with the numbering style
  $\{\lambda_1 \lambda_2 \mu_1 \mu_2\}$ shown explicitly.
  Then, we show the $R_{00cd},\; c,d=0...4$, correlation matrix (imaginary part equal zero)
  as calculated in the GPS frame, using eqs.~(\ref{R-matrix}) and (\ref{GPS-Born-amplitudes}),
  and after a (double) Wick rotation to JW2 $\tau$ rest frames.
  On the other hand, the result marked ``analytical'' is obtained from eq.~(\ref{R-matrix-JW}),
  also in the JW2 frames.
  The difference of the two results in the JW2 frame is shown to be precisely zero.
}
\label{fig1}
\end{figure}

\section{\normalsize GPS at work, $\tau$-pair spin correlations}
In the following we shall show how to apply the GPS rules
in order to find out the Wick rotation \cite{Wick:1962} responsible for change of 
spin quantization axes between
the JW and KS conventions for fermion spinors,
for the spin amplitudes of the $e^-e^+\to \tau^-\tau^+$ process.

Let us  use the variant of JW helicity states defined
in ref. \cite{Wick:1962} and refer to it as JW2.
The JW2 spin states are defined (quantized) 
in the $\tau^-$ rest frame with axes:
$\widehat{z} = -\widehat{\tau^+}$ and 
$\widehat{y} =  \widehat{\tau^+}\times \widehat{e^-}/ |\widehat{\tau^+}\times \widehat{e^-}|$,
and in the $\tau^+$ rest frame with axes:
$\widehat{z} = -\widehat{\tau^-}$ and 
$\widehat{y} = -\widehat{\tau^-}\times \widehat{e^+}/ |\widehat{\tau^-}\times \widehat{e^+}|$.
We denote as $\widehat{f}$ the unit three-vector parallel
to the momentum of a fermion $f$.
The other (third) axis is always picked in such a way
that the right-handed reference frame 
$(\widehat{x},\widehat{y},\widehat{z})$ is formed.
With the above choice of quantization axes the double spin
density matrix of the $\tau$ pair
(see formula (2.6) of ref.~\cite{jadach_was:1984},
used also in the KORALB Monte Carlo \cite{KORALB:1985}) 
takes the form:
\begin{equation}
  \label{R-matrix-JW}
 \{R^{JW2}_{00cd}\} =
  {\cal N}\;
  \begin{bmatrix} 
    1+c^2+M^2s^2 &  0           &  0             &  0            \\ 
       0         & (1+M^2)s^2   &  0             &  2Mcs        \\ 
       0         &  0           &  (1-M^2)s^2    &  0            \\ 
       0         & -2Mcs        &  0             &  -1-c^2+M^2s^2
  \end{bmatrix},
\end{equation}
where 
$c=\cos\theta$, $s=\sin\theta$, 
$\theta$ is the scattering angle between $e^-$ and $\tau^-$,
$M= 2m_\tau/s^{1/2}$ and  ${\cal N}=  1/( 1+c^2 +M^2s^2)$.
In order to simplify the discussion we include only  $s$-channel photon exchange 
(the parity-violating $Z$ exchange is neglected).
We also neglected the electron mass.
To be more precise, in ref.~\cite{jadach_was:1984} the spin quantization 
axes of $\tau^\pm$ are slightly different from JW2; let us call them the KB choice.
The KB spin states are defined (quantized) 
in the $\tau^-$ rest frame with axes 
$\widehat{z} =  \widehat{\tau^+}$ and 
$\widehat{x} =  \widehat{\tau^+}\times \widehat{e^-}/ |\widehat{\tau^+}\times \widehat{e^-}|$
and in the $\tau^+$ rest frame with axes
$\widehat{z} = -\widehat{\tau^-}$ and 
$\widehat{x} =  \widehat{\tau^-}\times \widehat{e^+}/ |\widehat{\tau^-}\times \widehat{e^+}|$.
The transition from KB to JW2 costs additional trivial Wick
rotation $R_3(\pi/2)$ for $\tau^+$  and $R_2(\pi)R_3(\pi/2)$ for $\tau^-$.
(These rotations are merely permutations and reflection of axes.)
This is what was actually done in order to get  (analytically)
our eq.~(\ref{R-matrix-JW}) from eq.~(2.6) of ref.~\cite{jadach_was:1984}.

On the other hand, we calculate spin amplitudes for the same process with
the basic KS spinors of eq.~(\ref{def-massive2}). Using the Chisholm identity 
(treating electron spinors as massless) and the inner product of eq.~(\ref{inner-massive}),
we obtain the following result:
\begin{equation}
\label{GPS-Born-amplitudes}
  \begin{split}
    {\cal M}^{\rm prod,GPS}_{\lambda_1 \lambda_2 \mu_1 \mu_2}
    = C\; \delta_{\lambda_1, -\lambda_2}
  \big[\; &\bar{u}(q_1, \mu_1)      u(p_1, \lambda_1)\;
           \bar{v}(p_2, \lambda_2)  v(q_2, \mu_2)\\
         +&\bar{u}(q_1, \mu_1)      v(p_2,-\lambda_2)\;
           \bar{u}(p_1,-\lambda_1)  v(q_2, \mu_2)\big]\\
     = C\;  \delta_{\lambda_1, -\lambda_2}
  \big[\; &S(q_1, m_\tau, \mu_1,      p_1,   0,    \lambda_1)\;
           S(p_2,   0,   -\lambda_2,  q_2,-m_\tau,-\mu_2 )\\
         +&S(q_1, m_\tau, \mu_1,      p_2,   0,    \lambda_2)\;
           S(p_1,   0,   -\lambda_1,  q_2,-m_\tau,-\mu_2)\big],
  \end{split}
\end{equation}
where $C$ is an overall normalization constant, unimportant for our discussion.
As compared to the previous result,
we lack insight into the analytical structure 
of the above formula;
it is simply meant to be evaluated numerically.
We are also unable to show analytical expressions for $R^{GPS}_{00cd}$ --
the translation from ${\cal M}^{\rm prod,GPS}_{\lambda_1 \lambda_2 \mu_1 \mu_2}$
to $R^{GPS}_{abcd}$ using eq.~(\ref{R-matrix}) has to be done numerically.
In view of the nice simplicity of the JW/KB helicity amplitudes
(see eg. eq. (2.3) of ref. \cite{jadach_was:1984}),
one may wonder why bother about all the complications of the KS/GPS scheme.
The answer is well known:
the KS spinors (now with GPS supplement) will be
very powerful and useful once we consider multiple photon emission from fermions.
In this case, the Jacob-Wick helicities are no longer so simple and useful,
as they were really invented for the Born amplitudes.

The following exercise would demonstrate that our GPS rules really work: 
If we calculate $R^{GPS}_{00cd}$ numerically
and if we really know precisely (from the GPS rules) in which $\tau^\pm$ reference
frames $R^{GPS}_{00cd}$ is defined, 
then we should be able by the appropriate two
non-trivial Wick rotations to get exactly the very simple $R^{JW2}_{00cd}$
of eq.~(\ref{R-matrix-JW}).
The two Wick rotations are in this case just ordinary rotations in two $\tau$ rest frames,
corresponding to transformations from the GPS to the JW2 rest frames.
In the following, see fig.~1, we show the numerical results, demonstrating precisely
the above operation done numerically.
Indeed, we see that due to the appropriate Wick rotations
the $R$-matrices become identical.
This demonstrates clearly that in fact our GPS supplement to the KS spinor
method works in the practical application.
The above exercise is just a small warm-up before
the forthcoming paper \cite{CEEX}, in which
we apply the GPS rules to a full-scale calculation with
multiple photon emissions (with exponentiation) and with decaying fermions.

\section{\normalsize Conclusions}
In the present paper we have established a method to control
the complex phase and spin quantization reference frame
for the Weyl spinor technique (Kleiss-Stirling type)
for massive and massless spinors.
The method applies to multiphoton emission. 
The present work opens the way to extending the work of 
refs.~\cite{jadach_was:1984,KORALB:1985,tauola2.4}
(where the spin effects, photon emission and Monte Carlo 
event generation were successfully combined
for the first time)
to the case of the multiple photon emission.
This will be elaborated in a forthcoming publication \cite{CEEX}.

\vspace{2mm}
\noindent
{\bf Acknowledgements}\\
We thank the CERN Theory Division and all four LEP collaborations for support.
ZW acknowledges specially the support of the ETH L3 group during the final work
on the paper preparation.
Useful discussions with E.~Richter-W\c{a}s are warmly acknowledged.


\end{document}